\documentclass[traditabstract, twocolumns]{aa}

\usepackage{graphicx}
\usepackage{txfonts}
\usepackage{natbib}
\usepackage{xspace}
\usepackage{scalerel}
\usepackage{multirow}
\usepackage{color}
\usepackage{xcolor}
\usepackage{hyperref}
\bibpunct{(}{)}{;}{a}{}{,}

\newcommand{\pycs}{{\tt PyCS}\xspace}

\newcommand{\hc}{$H_0$}
\def\ksmpc{${\rm\, km\, s^{-1}\, Mpc^{-1}}$\xspace}
\newcommand{\obj}{PG~1115$+$080}
\newcommand{\tel}{ESO MPIA 2.2m telescope}
\newcommand{\lcdm}{$\mathrm{\Lambda CDM}$\xspace}

\definecolor{crimson}{HTML}{DC143C}
\definecolor{royalblue}{HTML}{4169E1}
\definecolor{hotpink}{HTML}{FF69B4}
\definecolor{purple}{HTML}{800080}
\definecolor{midnightblue}{HTML}{191970}

\definecolor{indianred}{HTML}{CD5C5C}
\definecolor{seagreen}{HTML}{2E8B57}
\definecolor{steelblue}{HTML}{4682B4}
\definecolor{darkorange}{HTML}{FF8C00}
\definecolor{brown}{HTML}{A52A2A}

\begin{document}

\title{COSMOGRAIL XVII: Time delays for the quadruply imaged quasar 
\obj.}
\titlerunning{Time-delay measurements of \obj}

\author{
V. Bonvin \inst{\ref{epfl}} \and
J.~H.~H. Chan \inst{\ref{epfl}} \and
M. Millon \inst{\ref{epfl}} \and
K. Rojas \inst{\ref{valpo},\ref{lsst}} \and
F. Courbin \inst{\ref{epfl}} \and
G.~C.-F.~Chen \inst{\ref{ucdavis}} \and
C.~D. Fassnacht \inst{\ref{ucdavis}} \and
E. Paic \inst{\ref{epfl}} \and
M. Tewes \inst{\ref{bonn}} \and
D.C.-Y. Chao \inst{\ref{MPG}} \and
M. Chijani \inst{\ref{UNAB}} \and
D. Gilman \inst{\ref{ucla}} \and
K. Gilmore \inst{\ref{stanford}} \and
P. Williams \inst{\ref{ucla}} \and
E. Buckley-Geer \inst{\ref{fermilab}} \and
J. Frieman \inst{\ref{fermilab}, \ref{chicago}} \and
P.J. Marshall \inst{\ref{stanford}} \and
S.H. Suyu \inst{\ref{MPG}, \ref{TUM}, \ref{ASIAA}} \and
T. Treu \inst{\ref{ucla}} \and
A. Hempel \inst{\ref{UNAB}} \and 
S. Kim \inst{\ref{puc}, \ref{heidelberg}}\and
R. Lachaume \inst{\ref{puc}, \ref{heidelberg}}\and 
M. Rabus \inst{\ref{puc}, \ref{heidelberg}}\and
T. Anguita \inst{\ref{UNAB}, \ref{millenium}} \and
G. Meylan \inst{\ref{epfl}} \and
V. Motta \inst{\ref{valpo}} \and
P. Magain \inst{\ref{STAR}}
}

\institute{
Institute of Physics, Laboratory of Astrophysics, Ecole Polytechnique 
F\'ed\'erale de Lausanne (EPFL), Observatoire de Sauverny, 1290 Versoix, 
Switzerland \label{epfl}\goodbreak \and
Instituto de F\'isica y Astronom\'ia, Universidad de Valpara\'iso, Avda. 
Gran Breta\~na 1111, Playa Ancha, Valpara\'iso 2360102, Chile 
\label{valpo}\goodbreak \and
Departamento de Ciencias F\'isicas, Universidad Andres Bello Fernandez 
Concha 700, Las Condes, Santiago, Chile  \label{UNAB}\goodbreak \and
Centro de Astroingenier\'ia, Facultad de F\'isica, Pontificia Universidad 
Cat\'olica de Chile, Av. Vicu\~na Mackenna 4860, Macul 7820436, 
Santiago, Chile \label{puc}\goodbreak \and
Max-Planck-Institut f\"ur Astronomie, K\"onigstuhl 17, 69117 Heidelberg, 
Germany \label{heidelberg}\goodbreak \and
Department of Physics, University of California, Davis, CA 95616, USA 
\label{ucdavis}\goodbreak \and
Argelander-Institut f\"ur Astronomie, Auf dem H\"ugel 71, 53121, Bonn, 
Germany \label{bonn}\goodbreak \and
Fermi National Accelerator Laboratory, P.O. Box 500, Batavia, IL 60510, USA \label{fermilab}\goodbreak \and
Kavli Institute for Cosmological Physics, University of Chicago, Chicago, IL 60637, USA \label{chicago}\goodbreak \and
Kavli Institute for Particle Astrophysics and Cosmology, Stanford University, 452 Lomita Mall, Stanford, CA 94035, USA \label{stanford}\goodbreak \and
Max Planck Institute for Astrophysics, Karl-Schwarzschild-Strasse
1, D-85740 Garching, Germany \label{MPG}\goodbreak \and
Physik-Department, Technische Universit\"at M\"unchen, 
James-Franck-Stra\ss{}e~1, 85748 Garching, Germany \label{TUM}\goodbreak \and
Institute of Astronomy and Astrophysics, Academia Sinica, P.O.~Box 23-141, Taipei 10617, Taiwan \label{ASIAA}\goodbreak \and 
Department of Physics and Astronomy, University of California, Los Angeles, CA 90095, USA \label{ucla}\goodbreak \and
Millennium Institute of Astrophysics, Chile \label{millenium}\goodbreak \and
Space sciences, Technologies and Astrophysics Research (STAR) Institute, Universit\'e de Li\`ege, all\'ee du 6 Ao\^ut 17, 4000 Li\`ege, Belgium \label{STAR}\goodbreak \and LSSTC Data Science Fellow\label{lsst} \goodbreak
}

\date{\today}
\abstract{We present time-delay estimates for the quadruply imaged 
quasar \obj. Our resuls are based on almost 
daily observations for seven months at the \tel\ at La Silla 
Observatory, reaching a signal-to-noise ratio of 
about 1000 per quasar image. In addition, we re-analyse existing light curves from the literature that we complete with an additional three seasons of monitoring with the Mercator telescope at La Palma Observatory. When exploring the possible source of bias we consider the so-called microlensing time delay, a potential source of systematic error so far never directly accounted for in previous time-delay publications. In fifteen years of data on \obj, we find no strong evidence of microlensing time delay. Therefore not accounting for this effect, our time-delay estimates on the individual data sets are in good agreement with each other and with the literature. Combining the data sets, we obtain the most precise time-delay 
estimates to date on \obj, with $\Delta t(AB) = 8.3^{+1.5}_{-1.6}$ days (18.7\% precision), $\Delta t(AC) = 9.9^{+1.1}_{-1.1}$ days (11.1\%) and $\Delta t(BC) = 18.8^{+1.6}_{-1.6}$ days (8.5\%). Turning these time delays into cosmological 
constraints is done in a companion paper that makes use of ground-based Adaptive Optics (AO) with the Keck telescope.}


\maketitle

\section{Introduction}

\begin{figure*}[t!]
\centering
\includegraphics[width=0.95\linewidth]{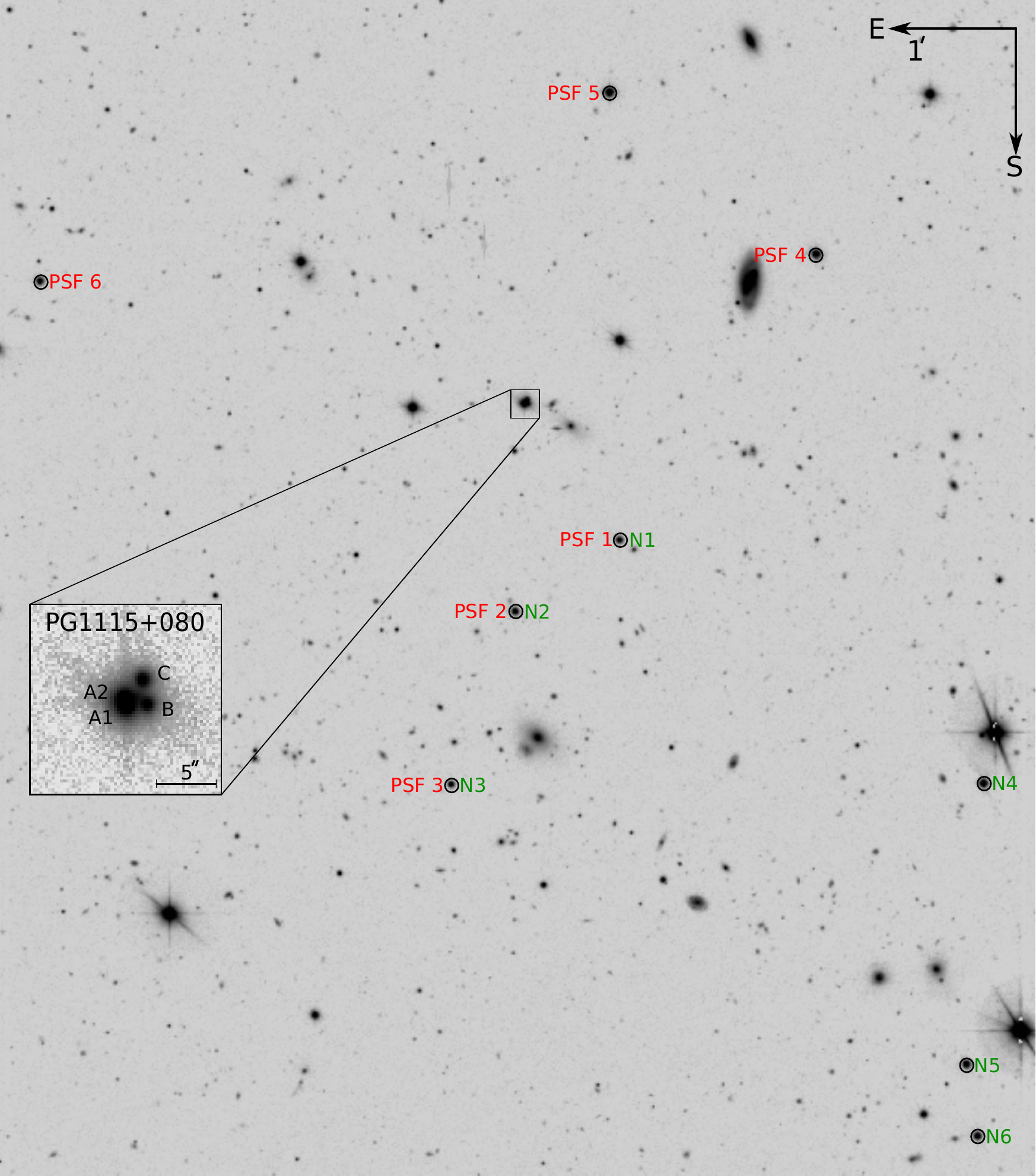}
\caption{Part of the field of view around \obj, as seen with the \tel\ at La Silla Observatory. The field is a stack of 92 
exposures with seeing $<$ 1.1" and ellipticity $<$ 0.12, for a total of 
$\sim$8.5 hours of exposure. The stars used for the modeling of 
the PSF are labeled PSF 1 to PSF 6, in red, and the stars used for the 
exposure to exposure normalisation are labeled N1 to N6, in green. The 
insert shows a single, 330-second exposure of the lens.}
\label{fig:fov}
\end{figure*}

The current cosmological paradigm is the standard cosmological model, also called flat-\lcdm. It assumes the 
presence of both dark energy in the form of a cosmological constant 
($\Lambda$) and 
cold dark matter (CDM), two components of unknown nature that have been 
puzzling scientists for decades. The flat-\lcdm model is determined by a set 
of \emph{cosmological parameters} whose values are jointly estimated in order for the model to match the 
observations.

The current expansion rate of the Universe, also called the Hubble constant or 
\hc, is one of these cosmological parameters whose value can be 
predicted in the flat-\lcdm model. Observations of the Cosmic 
Microwave Background (CMB) by the WMAP and Planck satellites put constraints on the flat-\lcdm model with values of the Hubble 
constant of \hc =70.0$\pm$2.2\ksmpc \citep{Bennett2013} and 
\hc =66.93$\pm$0.62\ksmpc \citep{Planck2016XLVI}. Large 
scale surveys are also helpful in that regard, finding values consistent with CMB predictions. Baryon Acoustic Oscillations yield in combination with CMB observations \hc=67.6$\pm$0.5 \citep{Alam2017}, and the Dark Energy Survey yields \hc=$67.2^{+1.2}_{-1.0}$ in combination with BAO but independently from CMB measurements \citep{DES2017}. 

In a complementary fashion, it is also possible to directly probe the Hubble constant in the local Universe by measuring the distance and recessional 
velocity of astrophysical objects of known intrinsic 
luminosity. These are labelled standard candles, or distance indicators \citep[see e.g.][]{Chavez2012, 
Freedman2012, Sorce2012, Beaton2016, Riess2016, Cao2017}. The currently most 
precise direct measurement of the Hubble constant comes from the so-called 
distance ladder technique, making use of cross-calibration of various 
distance indicators and yields a value of \hc=73.45$\pm$1.66\ksmpc 
\citep{Riess2018}, in tension with the flat-\lcdm prediction from the 
CMB observations and large-sky surveys.

Time-delay cosmography offers an independent approach to directly measure the Hubble constant. 
The original idea, postulated by \citet{Refsdal1964}, consists of 
measuring the time delay(s) between the luminosity variations of the 
multiple images of a strongly lensed source. Supernovae, due to their bright nature and variable luminosity were first considered as the ideal source but are however extremely rare \citep{Oguri2010}. Only 
two resolved occurrences have been observed to date, one located behind a cluser \citep[][labelled supernovae Refsdal]{Kelly2015, Kelly2016, Rodney2016} and the other located behind an isolated galaxy \citep{Goobar2017, More2017}. The 
discovery of the first lensed quasar \citep{Walsh1979}, whose occurences are much more numerous than supernovae, gave a huge boost to the field of time-delay cosmography. Over the years, time-delay cosmography has 
been refined up to the point that it yields nowadays one of the most 
precise measurement of \hc\ in the local Universe. In 
2016, the H0LiCOW\footnote{\url{www.h0licow.org}} collaboration 
\citep{Suyu2017} unveiled its measurement from a blind and thorough analysis 
of the gravitational lens HE\,0435-1223 \citep{Sluse2017, Rusu2017, 
Wong2017, Bonvin2017, Tihhonova2018}. Combined with previous efforts on two 
other lensed systems \citep{Suyu2010, Suyu2014}, it resulted in a value of 
\hc=$71.9^{+2.4}_{-3.0}$\ksmpc \citep{Bonvin2017}, in good agreement with the 
distance ladder but higher than the CMB predictions from the Planck 
satellite observations.

Whether this tension between the local and CMB measurements of \hc\ comes from unknown sources of errors, a statistical fluke or is the sign of new physics beyond flat-\lcdm is yet to be carefully examined. Concerning time-delay 
cosmography, increasing the overall precision and accuracy requires a larger sample of suitable strongly lensed systems. Recent years have seen 
the emergence of numerical techniques to find strong lenses candidates 
in surveys covering large portions of the sky 
\citep[e.g.][]{Joseph2014, Avestruz2017, Agnello2017, Petrillo2017, 
Lanusse2018} that result in the discovery new systems \citep{Lin2017, Agnello2017b, Schechter2017}. Once a new system is 
found, high-resolution imaging as well as time-delay measurements are mandatory for an in-depth analysis of the system. However, having to 
wait ten years for robust time-delay estimates is not viable, thus new 
monitoring strategies are currently being explored. 

In the framework of the 
COSMOGRAIL collaboration\footnote{\url{www.cosmograil.org}}, a 
high-cadence and high-precision monitoring campaign started in fall 2016 
on a daily basis at the \tel\ telescope at La Silla Observatory, 
in Chile (PI: Courbin). The first results were extremely encouraging, with a time 
delay measured at 1.8$\%$ precision between the two brightest images of 
DES J0408-5354 after only one season of monitoring \citep{Courbin2017}. 
In the present paper, we report the successful measurement of time 
delays 
on another lens system, \obj, after 7 months of monitoring. This 
measurement is combined 
with other time-delay estimates from previous monitoring campaigns and 
used in a companion paper to infer cosmological parameters (Chen et al. 
2018b, in prep).

\section{Observations and photometry}
\label{sec:data}

\obj\ is the second lensed quasar ever discovered 
\citep[$\alpha$(2000):\ 
11h18m17.00s; 
$\delta$(2000):~+07$^\circ$45'577" at redshift 
$z_s=1.722$][]{Weymann1980}. It has been identified as a quad in a fold 
configuration \citep{Hege1981}, whose two 
brightest images are separated by $\sim$0.5 arcseconds only. The 
redshift of the lens was determined more than a decade after the 
initial discovery, as 
$z_l=0.311$, independently by \citet{Kundic1997b} and \citet{Tonry1998}. 
The lens galaxy has been identified as being a member of a small group of 
galaxies \citep{Kundic1997b}. Infrared observations revealed 
the presence of an Einstein ring \citep{Impey1998}. The lens galaxy 
was later identified as elliptical 
\citep{Treu2002, Yoo2005}. The most recent determination of the astrometry of the 
system makes use of Hubble Space Telescope observations
\citep[see Table 1 of][]{Morgan2008}.

\subsection{High-cadence monitoring with the \tel}

The observational material for the present time-delay measurements 
consists of almost daily imaging data with the Wide Field Imager installed at the 
 \tel\ and taken between December 2016 and July 2017, called the WFI 
data set in the following. The full data set consists of 276 usable exposures of 330 
seconds each, for a total of $\sim$25 hours. The median seeing of the 
observations was 1.2$''$ and the median airmass was 1.31. Each WFI 
exposure consists of a 36' $\times$ 36$'$ field of view covered by eight 
CCDs, with a pixel size of 0.238$''$/pixel. The data reduction pipeline makes use of 
only one of the eight chips to ensure the stability of 
the night-to-night calibration. The exposures are all taken through the 
$\mathrm{ESO\ }BB\#R_c/162$ filter centered around $651.725\ nm$. 

The data 
reduction process follows the standard pipeline already in the 
most recent COSMOGRAIL publications \citep{Tewes2013b, Rathnakumar2013, 
Bonvin2017, Courbin2017}. It includes bias subtraction, flat 
fielding, sky removal with Sextractor \citep{Bertin1996}, fringe 
pattern removal as well as PSF reconstruction and source deconvolution using the 
MCS algorithm \citep{Magain1998, Cantale2016}. Figure 
\ref{fig:fov} presents a stack of the 92 exposures with seeing $<$ 
1.1" and ellipticity $<$ 0.12, as well as a single 330-second cutout 
of the lens seen with WFI.  The stars used for the 
PSF reconstruction are labeled PSF 1 to PSF 6 in red and the stars used 
for the exposure-to-exposure normalization are labeled N1 to N6 in 
green. Because the A1 and A2 images are separated by only a few tenths of arcseconds - too close for our deconvolution scheme to be properly resolved - their measured fluxes are merged together in a single component simply called A. The resulting light curves are presented in the top-left panel 
of Fig.~\ref{fig:lcs}.\footnote{The WFI reduced light curves will be available on the COSMOGRAIL website and CDS upon acceptance of the paper for publication.} 

\subsection{Previous datasets}

\begin{figure*}[t!]
\centering
\includegraphics[width=18cm]{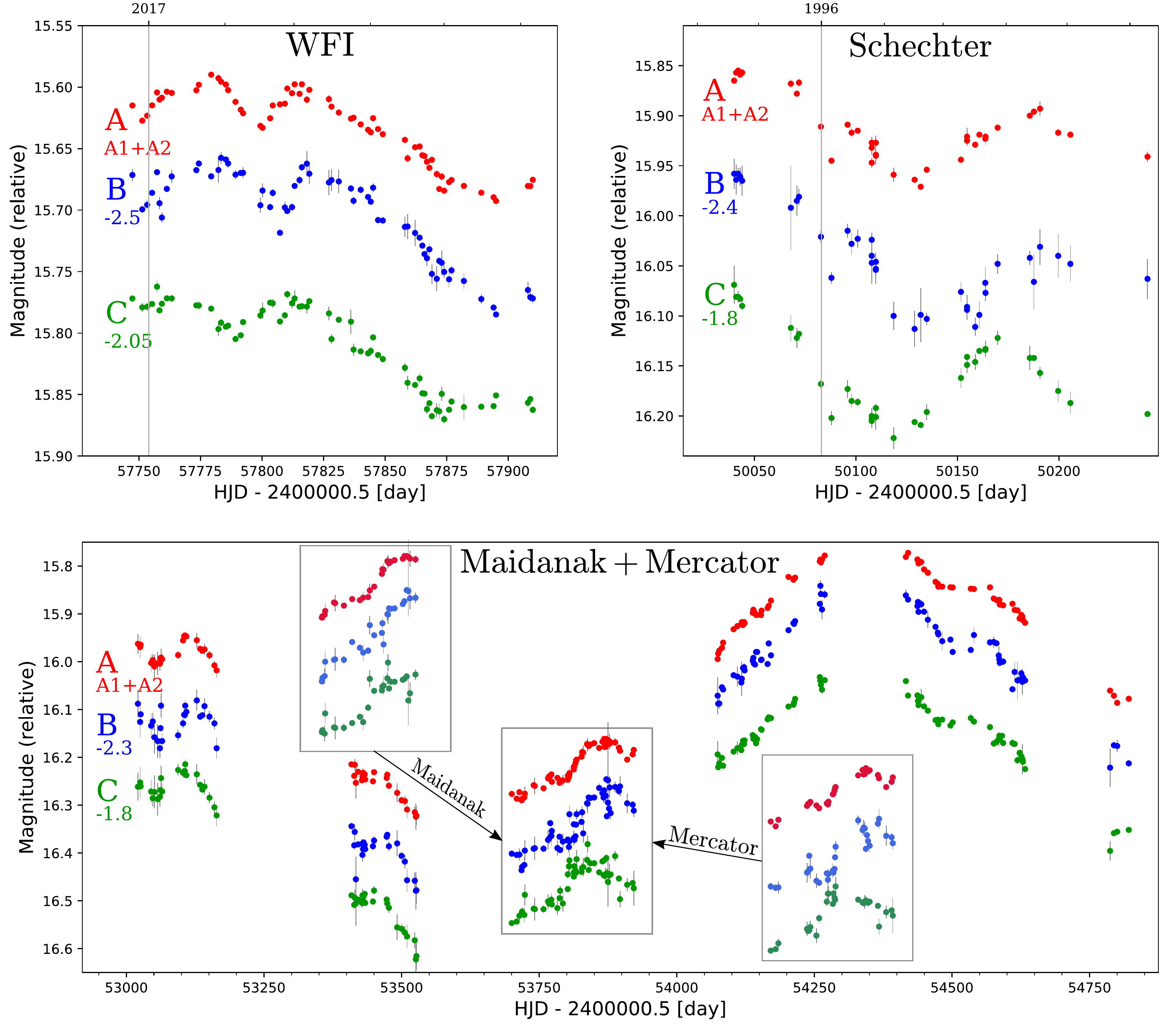}
\caption{Light curves for the three data sets used in this work. The B and C components of each data set have been shifted in magnitude for visual purposes. The A component corresponds to
the integrated flux of the unresolved A1 and A2 quasar images. The WFI and Mercator data are new, while the other observations in the Schechter and Maidanak data were published in \citet{Schechter1997} and \citet{Tsvetkova2010}, respectively (see Fig.2 of both papers). Maidanak and Mercator data overlap during the 2006 season only (see text for details). The inserts show the contribution to the 2006 season from both data sets.}
\label{fig:lcs}
\end{figure*}

In addition to the WFI data, we make use of the already reduced 
data obtained at the Maidanak telescope in Uzbekistan in the years 
2004-2006\footnote{Test data were acquired during the 2001-2003 
seasons, but are too sparsely sampled to bring any constraints on the 
time-delay measurements. We thus disregarded them in the present work.} 
\citep{Tsvetkova2010}. We complement these observations with three extra years of monitoring at the Mercator telescope between 2006 and 2009, whose observing cadence and photometric precision are comparable to the Maidanak data. The Mercator data reduction was done using the same pipeline as the WFI data, yet using different normalization and PSF stars due to different CCD size and defects (position of dead pixels and dead lines). The Mercator data set not being of sufficient quality to measure time delays on its own, it is merged with Maidanak into a single set (hereafter the Maidanak+Mercator data set) presented in the bottom panel of Fig.~\ref{fig:lcs}, where each Mercator light curve was independently shifted in magnitude in order to overlap with the 2006 season of its Maidanak counterpart. If the A light curves overlap very well, B and C show some discrepancy of unknown origin in the second part of the 2006 season, e.g. around MHJD=53880 days. Robustness checks performed when measuring the time delays showed that this discrepancy between the two instruments had no visible effect on the time-delay measurements.

Finally, we complement our analysis with data points from the Hiltner, WIYN, 
NOT and Du Pont telescopes acquired in 1996-1997 
and first presented in \citet{Schechter1997} (hereafter the Schechter data 
set - data courtesy of P. Schechter). The corresponding light curves are 
reproduced 
in the top-right panel of Fig.~\ref{fig:lcs}.

\section{Time delay measurement}
\label{sec:tds}

To estimate the time delays we use {\tt PyCS}\footnote{\pycs can be obtained from 
\url{http://www.cosmograil.org}}, a publicly available python toolbox developed by the COSMOGRAIL collaboration. \pycs was originally presented in \citet{Tewes2013a}. Since then, it has been continuously 
developed, applied to a rapidly growing number of data sets \citep[see e.g.][]{Tewes2013b, Eulaers2013, 
Rathnakumar2013, Bonvin2017, Courbin2017} and extensively tested in the 
scope of the Time Delay Challenge \citep{Liao2015, 
Bonvin2016}.

\subsection{\pycs formalism}
\label{sec:clockworks}

The formalism used in \pycs is presented in full details 
in \citet{Tewes2013a}, of which we summarize here the key aspects. 

A \emph{curve-shifting technique} designates a procedure that takes a set of light curves as input and yields the corresponding time-delay estimates with associated uncertainty. In this approach, a curve-shifting technique is defined by i) an \emph{estimator} that is an algorithm 
yielding the optimal point estimates of the time delay(s) in a set of light curves, ii) \emph{estimator parameters} that control the behaviour and convergence of the estimator and iii) an \emph{error estimation procedure} that assesses the robustness of the estimator. \pycs currently makes use of two estimators:

\begin{enumerate}

 \item The  \emph{free-knot 
splines estimator} makes use of splines, which are piece-wise 3rd order 
polynomials 
linked together by \emph{knots} at which the 2nd derivative is 
continuous. The estimator fits a spline to model the common intrinsic luminosity variations of the 
quasar and individual extrinsic splines to model the luminosity variations due to 
microlensing independently affecting each light curve. The 
overall variability is controlled by the \emph{initial knot step 
$\eta$} of the 
splines. The local variability of the splines is adapted to 
match the observed features by iteratively adjusting the position of 
the knots, coefficients of the polynomials and both time and 
magnitude shifts of the light curves, following the 
bounded-optimal-knot 
algorithm presented in \citet{Molinari2004}. 

 \item The \emph{regression 
difference estimator} independently fits regressions through each 
individual 
light curve using Gaussian procesess whose \emph{covariance function, amplitude, 
scale} and \emph{observation variance} can be adjusted. The regressions are then shifted in time 
and subtracted pair-wise. The amount of variability of the subtraction, i.e. a quantification of how ``flat'' the subtraction is,  is 
computed for each time shift. The minimum in variability corresponds to the 
optimal time shift of the estimator.

\end{enumerate}

The estimated mean value of the time delays is obtained by running the chosen estimator 
200 times on the data, each time from a different starting point, and 
taking the mean result. Note that this process is not a Monte-Carlo approach: only the initial conditions to the fit are changed between two runs, which are applied to the exact same data. A large dispersion of the measured values indicates that the estimator fails to converge for the given choice of estimator parameters.

The error estimation procedure used in \pycs consists of drawing sets of mock light 
curves from a generative model, based on the quasar intrinsic variability and 
individual slow extrinsic variability curves as modeled by the free-knot spline 
technique applied on the data. The residuals of the fit are used to compute the 
statistical properties of the correlated noise and any other signal not included in the fit. Therefore, each set of mock light curves has the 
same intrinsic and slow extrinsic variations, but a different realisation of the 
noise drawn with respect to a common set of statistical properties. In addition, ``true" 
time delays for each set are randomly chosen in an interval around the measured 
delays on the original data. Assuming that these sets of mock curves mimic 
plausible realisations of the observations, the errors on the time-delay estimates can be computed by 
comparing the result of the estimator applied on each set of mock to their true delays. 
Exploring a large range of true delays allows one to detect any ``lethargical'' 
behaviour in the estimator \citep{Rathnakumar2013} by binning the resultings 
errors according to the true delays of the mocks and checking if there is a systematic bias
evolving with the value of the true delays. In practice, we draw 1000 sets of mock light curves. The final errors consist of the worst systematic and random errors accross all bins, added in quadrature. Provided there is no apparent lethargical behaviour and that the systematic part of the errors is smaller than the random one, the estimated mean values and associated errors can be associated to Gaussian probability distributions. These probability distribution functions will be used later when combining various sets of estimates together.

A complete analysis of a data set thus requires one to choose i) an estimator, ii) 
the parameters of this estimator and iii) the method parameters of the 
free-knot splines estimator used in the generative model of mock curves. Together, 
these three criteria define a curve-shifting technique. Obviously, not 
all possible combinations of estimators and parameters are wise. For 
example, choosing a too small or too large initial knot step when fitting 
free-knot splines can lead to over or under fitting of the data, 
respectively assuming unphysical variations or missing information in the 
data. However, most choices of estimator parameters 
leading to a bad fit of the data result in larger dispersions when computing the means and/or the errors \citep[see][for an illustration]{Tewes2013b}. Ultimately, in this 
data-driven approach the preferred curve-shifting technique is the one 
yielding the most precise time-delay estimates. It is up to the \pycs 
user to assess the robustness of the curve-shifting technique used, by 
ensuring that slight modifications of the estimator parameters do not 
significantly affect the final results. A visual description of the 
pipeline detailed here can be found in Fig.~2.6 of \citet{BonvinThesis}.

\subsection{Application to the individual data sets}

\begin{figure*}[h!]
\centering

   \includegraphics[width=0.94\textwidth]{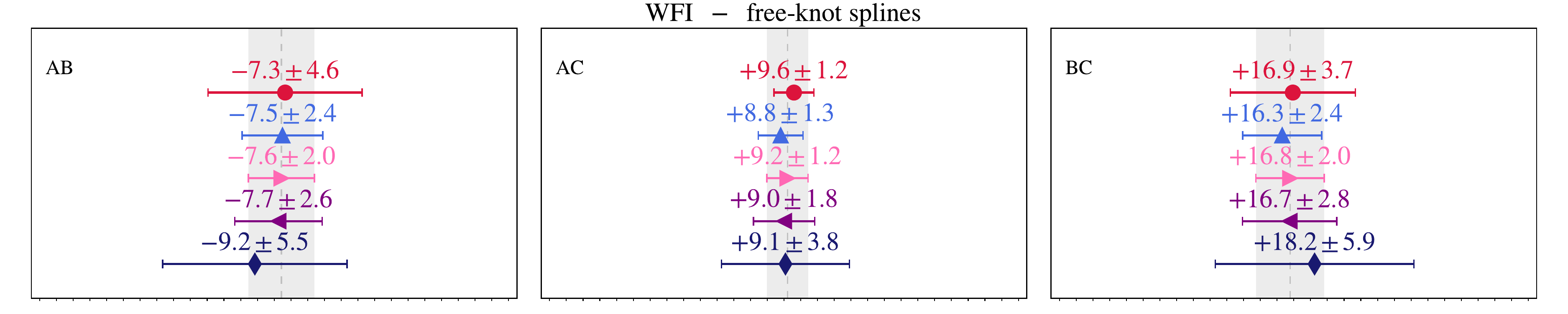}

   \includegraphics[width=0.94\textwidth]{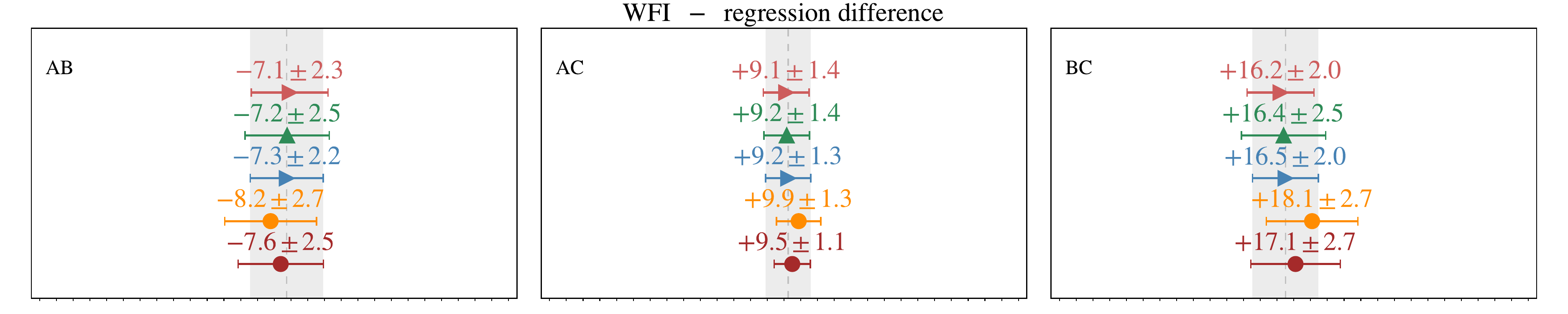}
     
   \includegraphics[width=0.94\textwidth]{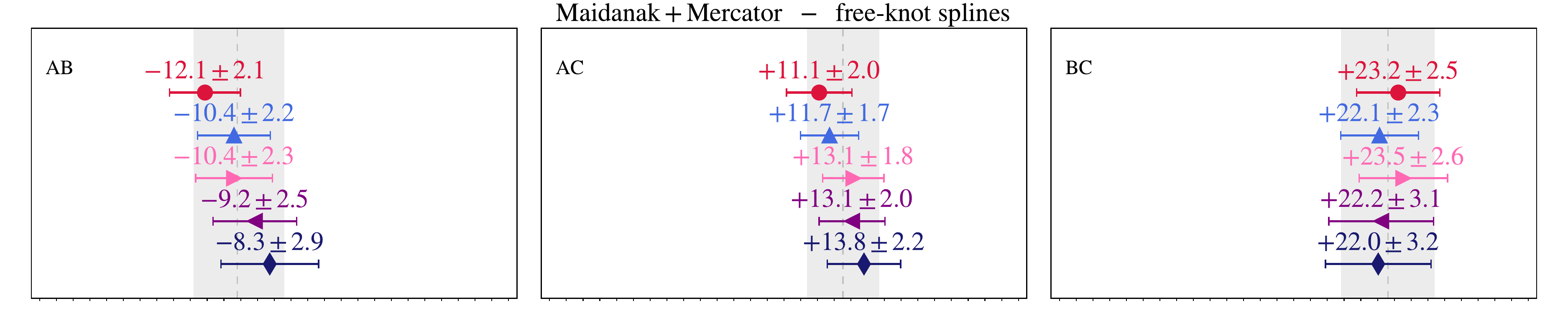}
         
   \includegraphics[width=0.94\textwidth]{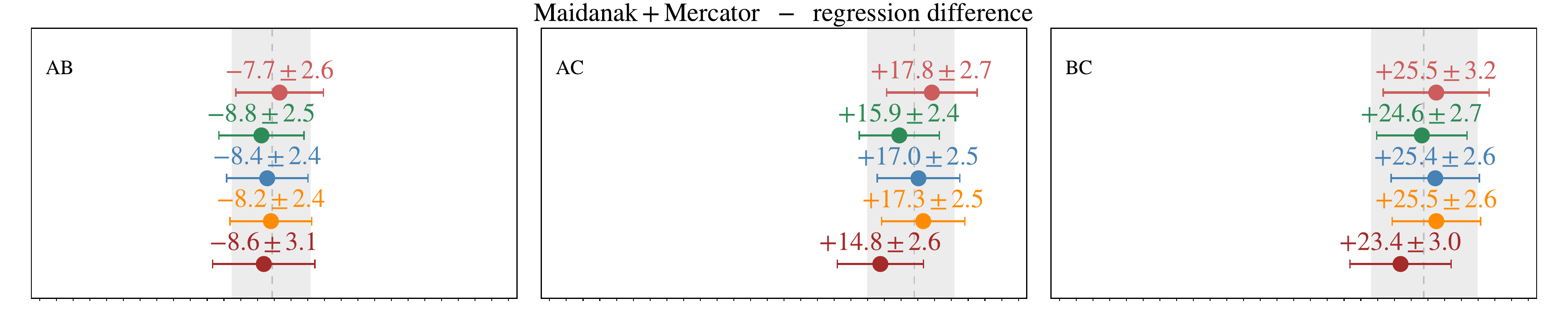}
   
   \includegraphics[width=0.94\textwidth]{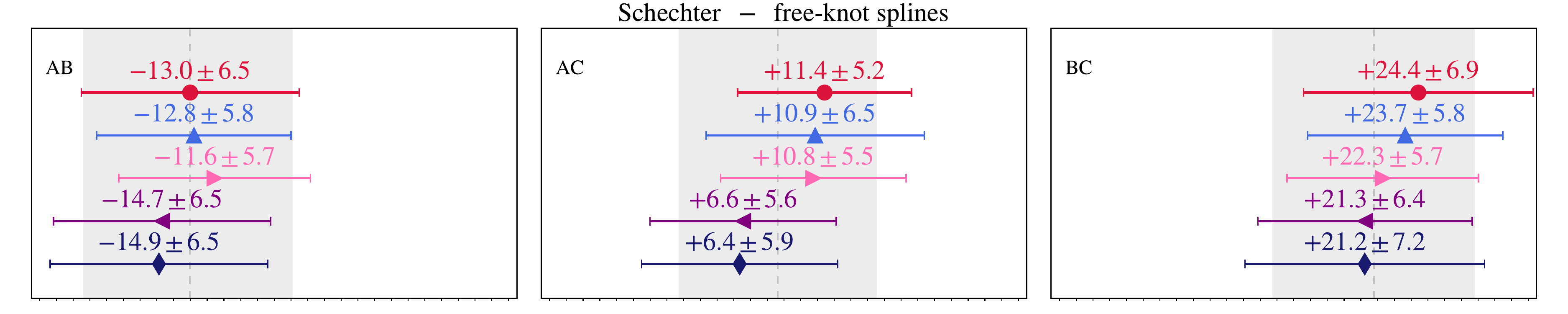}
         
   \includegraphics[width=0.94\textwidth]{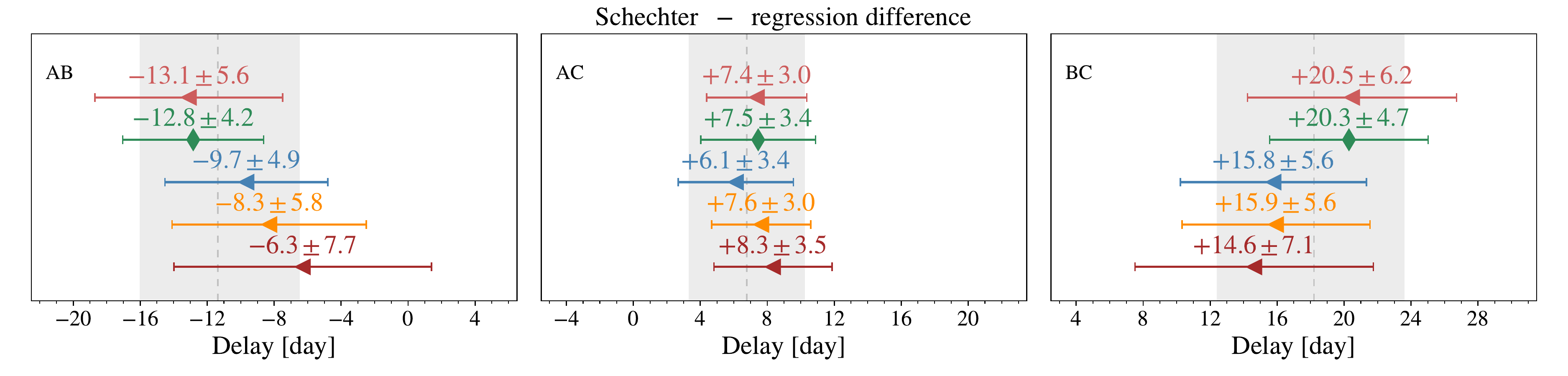}   
   
\caption{Time delays estimates and uncertainties (including both the statistical and systematic contributions) between the three pairs of light curves of \obj. Each column corresponds to a given pair of light curves, indicated in the top-left corner of each panel. Each row corresponds to a series, i.e. groups of time-delay estimates applied on given data set and curve-shifting technique, the name of which is indicated above the central panel. The estimator parameters corresponding to each group of time-delay estimates are indicated in Tab.~\ref{tab:tabparams}. For each two consecutive rows, i.e. time-delay estimates from the same data set, the symbols correspond to the generative model used when drawing the mock light curves. The shaded region in each panel indicates the combined time-delay estimates for $\tau_{\rm{thresh}}$=0.5 (see text for details).}
\label{fig:alldelays}
\end{figure*}


\begin{table*}[!ht]
\centering
  \caption{List of the estimator parameters used to compute the time-delay estimates presented in Fig.~\ref{fig:alldelays}. For the free-knot splines technique, $\eta$ corresponds to the initial knot spacing of the intrinsic spline, $\eta_{\mathrm{ml}}$ to the initial knot spacing of the extrinsic microlensing splines and $\eta_{\mathrm{ml}}$ pos. to constraints on the position of such knots. ``-'' for $\eta_{\mathrm{ml}}$ indicates that the microlensing splines have a single knot, regardless of their length. For the regression difference technique the parameters $\nu$ (smoothness degree), A (amplitude in magnitudes), scale (length scale in days) and errscale (observation variance in days) refer to the Mat\'ern covariance function used in the Gaussian process regression implementation of the \texttt{pymc.gp} module \citep[see][]{Tewes2013a, Patil2010}. In the rightmost column (brown symbol, marked with an *), the Mat\'ern covariance function is replaced by a power-law covariance function and $\nu$ indicates the power-law index used.}

   \begin{tabular}{l| c  c  c  c  c  c | c  c  c  c  c  c }

    \multicolumn{1}{c}{} & \multicolumn{6}{c|}{free-knot splines} & \multicolumn{6}{|c}{regression difference} \\ \hline\hline
    
    \multirow{5}{*}{WFI} & &
    \raisebox{-1.5pt}{\textcolor{crimson}{\scaleobj{1.6}{\bullet}}} &
    \raisebox{-0.75pt}{\textcolor{royalblue}{\scaleobj{1.3}{\blacktriangle}}} &
    \raisebox{-1.25pt}{\textcolor{hotpink}{\scaleobj{1.3}{\blacktriangleright}}} &
    \raisebox{-1.25pt}{\textcolor{purple}{\scaleobj{1.3}{\blacktriangleleft}}} & 
    \raisebox{-1.25pt}{\textcolor{midnightblue}{\scaleobj{1.3}{\blacklozenge}}} &  & \raisebox{-1.25pt}{\textcolor{indianred}{\scaleobj{1.3}{\blacktriangleright}}} & 
    \raisebox{-1.25pt}{\textcolor{seagreen}{\scaleobj{1.3}{\blacktriangle}}} &
    \raisebox{-1.25pt}{\textcolor{steelblue}{\scaleobj{1.3}{\blacktriangleright}}} & 
    \raisebox{-1.25pt}{\textcolor{darkorange}{\scaleobj{1.6}{\bullet}}} & 
    \raisebox{-1.25pt}{\textcolor{brown}{\scaleobj{1.6}{\bullet}}}* \\ 

    & $\eta$ & 15 & 20 & 30 & 40 & 50 &
    $\nu$ & 1.7 & 1.8 & 1.5 & 1.3 & 1.9 \\ 
    
    & \multirow{2}{*}{$\eta_{\mathrm{ml}}$} & \multicolumn{5}{c|}{\multirow{2}{*}{-}} & 
    A & 0.5 & 0.6 & 0.4 & 0.3 & 0.7 \\ 
        
    & & \multicolumn{5}{c|}{} & scale & 200 & 150 & 250 & 150 & 250 \\ 
    
    & $\eta_{\mathrm{ml}}$ pos. & \multicolumn{5}{c|}{unique knot fixed at center} &
    errscale & 20 & 15 & 25 & 10 & 25  \\ \hline

    \multirow{5}{*}{Maidanak + Mercator} & &
    \raisebox{-1.5pt}{\textcolor{crimson}{\scaleobj{1.6}{\bullet}}} &
    \raisebox{-0.75pt}{\textcolor{royalblue}{\scaleobj{1.3}{\blacktriangle}}} &
    \raisebox{-1.25pt}{\textcolor{hotpink}{\scaleobj{1.3}{\blacktriangleright}}} &
    \raisebox{-1.25pt}{\textcolor{purple}{\scaleobj{1.3}{\blacktriangleleft}}} & 
    \raisebox{-1.25pt}{\textcolor{midnightblue}{\scaleobj{1.3}{\blacklozenge}}} &  & \raisebox{-1.25pt}{\textcolor{indianred}{\scaleobj{1.6}{\bullet}}} & 
    \raisebox{-1.25pt}{\textcolor{seagreen}{\scaleobj{1.6}{\bullet}}} &
    \raisebox{-1.25pt}{\textcolor{steelblue}{\scaleobj{1.6}{\bullet}}} & 
    \raisebox{-1.25pt}{\textcolor{darkorange}{\scaleobj{1.6}{\bullet}}} & 
    \raisebox{-1.25pt}{\textcolor{brown}{\scaleobj{1.6}{\bullet}}}* \\ 

    & $\eta$ & 20 & 30 & 40 & 50 & 60 &
    $\nu$ & 2.2 & 1.8 & 1.9 & 1.9 & 1.8 \\ 
    
    & \multirow{2}{*}{$\eta_{\mathrm{ml}}$} & \multicolumn{5}{c|}{\multirow{2}{*}{200}} & 
    A & 0.5 & 0.7 & 0.6 & 0.4 & 0.7 \\

    & & \multicolumn{5}{c|}{} & scale & 200 & 200 & 200 & 200 & 250 \\  
 
    & $\eta_{\mathrm{ml}}$ pos. & \multicolumn{5}{c|}{min. 100 days btw knots} &
    errscale & 25 & 25 & 20 & 10 & 25 \\ \hline

    \multirow{5}{*}{Schechter} & &
    \raisebox{-1.5pt}{\textcolor{crimson}{\scaleobj{1.6}{\bullet}}} &
    \raisebox{-0.75pt}{\textcolor{royalblue}{\scaleobj{1.3}{\blacktriangle}}} &
    \raisebox{-1.25pt}{\textcolor{hotpink}{\scaleobj{1.3}{\blacktriangleright}}} &
    \raisebox{-1.25pt}{\textcolor{purple}{\scaleobj{1.3}{\blacktriangleleft}}} & 
    \raisebox{-1.25pt}{\textcolor{midnightblue}{\scaleobj{1.3}{\blacklozenge}}} &  & \raisebox{-1.25pt}{\textcolor{indianred}{\scaleobj{1.3}{\blacktriangleleft}}} & 
    \raisebox{-1.25pt}{\textcolor{seagreen}{\scaleobj{1.3}{\blacklozenge}}} &
    \raisebox{-1.25pt}{\textcolor{steelblue}{\scaleobj{1.3}{\blacktriangleleft}}} & 
    \raisebox{-1.25pt}{\textcolor{darkorange}{\scaleobj{1.3}{\blacktriangleleft}}} & 
    \raisebox{-1.25pt}{\textcolor{brown}{\scaleobj{1.3}{\blacktriangleleft}}}* \\ 

    & $\eta$ & 40 & 50 & 60 & 70 & 80 &
    $\nu$ & 2.2 & 1.8 & 1.5 & 1.2 & 1.6 \\
    
    & \multirow{2}{*}{$\eta_{\mathrm{ml}}$} & \multicolumn{5}{c|}{\multirow{2}{*}{-}} & 
    A & 0.5 & 0.7 & 0.4 & 0.3 & 0.2 \\ 
    
    & & \multicolumn{5}{c|}{} & scale & 250 & 250 & 250 & 350 & 350 \\ 

    & $\eta_{\mathrm{ml}}$ pos. & \multicolumn{5}{c|}{unique knot fixed at center} &
    errscale & 85 & 55 & 25 & 65 & 55 \\ \hline\hline    
    
  \end{tabular}
   
\label{tab:tabparams}  
\end{table*}

The three data sets presented in Sec.\ref{sec:data} can 
in principle be handled 
by \pycs together as a two-decade long monitoring campaign, with large gaps of many years 
in between. We choose however not to proceed this way, as the data sets have a different sampling cadence and 
photometric accuracy and thus are sensitive to features of different 
timescale. Analyzing them together requires the choice of a 
given knot step for the initial splines fit that is at the core of the 
generative model. As stated above, the knot step is a key parameter of 
the spline estimator. Forcing a single knot step for a common fit will average out the knot repartition over the three campaigns, 
\emph{de facto} over- or under-fitting some of the most shallow/sharp 
intrinsic variations features in the data.
Since the three monitoring campaigns are i) separated by gaps of six 
and eight years, ii) shorter than these gaps and iii) 
displaying no clear signs of decade-long correlated variability, we safely 
conclude that we can treat them independently and combine the resulting 
time delays a posteriori.

In addition to the \emph{curve-shifting technique} and associated definitions presented in Sec.~\ref{sec:clockworks}, we use the following terminology:

\begin{itemize}

 \item A \emph{data set} $\vec{D}$ refers to either the WFI, Maidanak+Mercator of Schechter monitoring campaigns.
 
 \item A \emph{time-delay estimate} $\vec{E} = \Delta t^{+\delta t_+}_{-\delta t_-}$ is a measurement of the mean and associated upper and lower errors between a given pair of light curves of a given data set. It corresponds to each single measurement in Fig.~\ref{fig:alldelays}.
 
 \item A \emph{group} of time delay estimates $\vec{G}=[\vec{E_{AB}}, \vec{E_{AC}}, \vec{E_{BC}}]$ represents the time delays between all pairs $j$ of light curves of the lensed quasar, measured by a given curve shifting technique applied on a given data set. It corresponds to any three points of the same color in each row of Fig.~\ref{fig:alldelays}.
 
 \item A \emph{series} of time delay estimates $\vec{S}=[\vec{G_1},...\vec{G_i},...\vec{G_N}]$ for $i \in N$ is an ensemble of groups of time delay estimates that share the same data set and estimator. A series is typically obtained by varying the estimator parameters and/or error estimation procedure of a curve-shifting technique. It corresponds to each row of Fig.~\ref{fig:alldelays}, where $N=5$ in this case. 
 
\end{itemize}

We process each data set the same way. First, we apply the 
\emph{free-knot splines technique}, consisting of the free-knot spline estimator 
applied on the data and mock curves analysis as well as in the mock curves generative model. While exploring various choices of estimator parameters, we choose to use the same 
parameters when fitting the data, the mocks and in the generative model in order to limit the 
number of possible configurations to consider. 
Similarly, we focus on only one type of slow microlensing modeling. For the 
shorter data sets (WFI and Schechter), we follow \citet{Courbin2017} by using 
extrinsic splines with a single knot whose position is fixed on the time axis at the 
middle of the light curves. For the longer data set (Maidanak+Mercator) we use splines 
with roughly one knot per season, whose position is free to vary during the iterative 
fitting process, up to a minimal distance of 100 days between the knots. Using this representation for microlensing, we are left with only one 
estimator parameter to vary, which is the initial knot step $\eta$ of the  
spline used to represent the intrinsic variations of the quasar. Eyeballing the fitting of 
the original data gives us an $\eta$ to start with, and other $\eta$ values are 
explored around this initial guess. As stated earlier, an inappropriate choice of $\eta$ yields time-delay estimates with larger 
error bars, thus giving us upper and lower limits for $\eta$. The resulting series of time-delay estimates, obtained by using five different $\eta$ for each data set can be seen in every second row of Fig.~\ref{fig:alldelays}, with ``free-knot splines'' in the subtitle. 

Second, we apply the \emph{regression difference technique} consisting 
of the regression difference estimator used in the data and mock curves 
analysis, while still using the free-knot spline estimator in the generative model. Here, the choice of the regression difference estimator parameters to fit 
the data and mock curves is completely independent from the choice of free-knot splines estimator parameters used in the generative model. We first choose five different plausible combinations of 
regression 
difference estimator parameters. For simplicity, we decide to use the 
same generative models as for the free-knot splines technique. Therefore, to each of the 
five combinations of estimator parameters correspond five possible generative models, each of which influences only the precision of 
the resulting group of time-delay estimates. For each choice of regression difference estimator 
parameters, there is one generative model that yields the most precise group of time-delay  
estimates. In order to assess which group is the most precise, we define the \emph{relative precision} of a series of time-delay estimates: for each group $i$ in the series, the relative precision reads as

\begin{equation}\label{eq:precision}
 P_i = \sum\limits_{j}\frac{\delta t_{i,j,+} + \delta t_{i,j,-}}{2\Delta t_{j}}, \ \ \Delta t_j = \frac{\sum\limits_{i}\left(\delta t_{i,j,+} + \delta t_{i,j,-}\right)\Delta t_{i, j}}{\sum\limits_{i}\left(\delta t_{i,j,+} + \delta t_{i,j,-}\right)},     
\end{equation}

\noindent where we sum over the $j$ time delay estimates of each group, and where $\Delta t_j$ is the mean of the individual $j$ delays over the $i$ groups of the series. Note that we compute $\Delta t_j$ using the unweighted mean of the $\Delta t_{i,j}$ for simplicity. The results for various choices of estimator parameters are presented in each second row of Fig. \ref{fig:alldelays}, with ``regression difference'' mentioned in the subtitle. Each group represents one choice of regression difference estimator parameters, and the symbols indicate which generative model from the corresponding free-knot spline row above has been used. 
 
\section{Towards a single group of time-delay estimates} 

A latent question of \pycs concerns the combination of multiple groups of time-delay estimates obtained with different curve-shifting techniques
in order to get a definitive measurement. By construction, a given 
curve-shifting technique has always one set of estimator parameters for 
which the best precision is achieved. The expected behaviour when 
varying the estimator parameters is 
that it impacts mostly the precision while marginally affecting the mean of the
measured time delays. Such behaviour has always been observed in the previous 
COSMOGRAIL work and was part of the usual robustness 
checks, as mentioned earlier. However, the present 
case is the first time where light curves not obtained with the 
COSMOGRAIL reduction pipeline are thoroughly analyzed with \pycs. As observed in Fig.~\ref{fig:alldelays}, the measured mean time delays for the Maidanak+Mercator and Schechter data sets shift with the choice of estimator parameters. 
Furthermore, the best estimates from two different curve-shifting 
techniques are not necessarily in excellent agreement. For example,
\citet{Tewes2013b} and \citet{Bonvin2017} each present time-delay estimates from 
both the regression difference and free-knot splines techniques 
but pick the most precise as the absolute reference, whereas the 
two techniques agree at the $\sim$1-sigma level. In this section, we first combine the groups of time-delay estimates per data set, marginalizing over the possible choice of estimator parameters and curve-shifting techniques weighted by their individual precision. We then discuss the combination of the estimates from the three data sets together and propose two possible results.

\subsection{Combining various curve-shifting techniques}

In order to combine various groups of time-delay estimates measured on the same data set, we make use of the precision $P$ defined in Eq.~\ref{eq:precision} but also of the \emph{tension} between two groups. For two time-delay estimates $\vec{E_A} = A_{-a_-}^{+a_+}$ and $\vec{E_B} = B_{-b_-}^{+b_+}$ with $A>B$, the tension in $\sigma$ units is defined as 

\begin{equation}
\tau(E_A, E_B)=(A-B)/\sqrt{a_{-}^2 + b_{+}^2}.
\end{equation}

For reference, two Gaussian 
distributions overlapping at their respective 1$\sigma$ (2$\sigma$) points thus 
have a tension of $\tau=\sim1.4\sigma\,(2\sigma)$. Therefore, the tension between two groups $\vec{G_1}$ and $\vec{G_2}$ is:

\begin{equation}\label{eq:maxtension}
 \tau_{\vec{G_1}, \vec{G_2}} = \max\limits_{j}(\tau(\vec{E_{1, j}}, \vec{E_{2, j}})), 
\end{equation}

\noindent i.e. the maximum tension between the time-delay estimates from corresponding pairs of light curves. In order to combine the time-delay estimates together, we proceed in the following way: for each serie of time-delay estimates sharing the same data set and estimator (i.e. each row of Fig.~\ref{fig:alldelays}), we first pick the most precise group in the series as our reference $\vec{G_{ref}}$. In previous COSMOGRAIL publications, this reference would have been our definitive group of time-delay estimates for the considered curve-shifting technique, but in the present case it is rather a starting point that we might or not combine with other groups. To do so, we compute the tension between each group $i$ in the series and the reference group $\tau_{i, ref}$. If the tension exceeds a certain threshold $\tau_{\rm{thresh}}$, the corresponding group is flagged. We then pick the most precise of the flagged groups and combine it with the reference group by marginalizing over the respective time-delay estimates probability distributions. This creates a new reference group. We then repeat the procedure above with the remaining groups, until there is none exceeding the tension threshold $\tau_{\rm{thresh}}$. The reference group is then considered as the final group of time-delay estimates for the considered curve-shifting technique and data set. The combined reference estimates are displayed as gray shaded regions in each panel of Fig.~\ref{fig:alldelays} for $\tau_{\rm{thresh}}=0.5\sigma$. Note that the combined estimates do not follow a Gaussian probability distribution anymore; in such cases, we take as the mean and 1$\sigma$ error bars the 50th, 16th and 84th percentiles of the distribution, respectively.

The results of the two estimators can then be combined together. The free-knot spline technique and regression difference technique, although fundamentally different in their conception cannot be considered as independent estimates when applied to the same data set. Thus, for each data set, the two corresponding sets of time-delay estimates (shaded gray regions in Fig.~\ref{fig:alldelays}) are considered as equiprobable distributions that are marginalized (i.e., the probability distributions are summed) over to yield a final group of estimates per data set. The combined group of time-delay estimates are presented in Fig.~\ref{fig:finalests}, labelled ``PyCS-WFI'', ``PyCS-Maidanak+Mercator'' and ``PyCS-Schechter''. They can be compared to time-delay estimates from the literature that use the same data sets. The Schechter data set has been analyzed by \citet{Schechter1997, Barkana1997, Pelt1998}. The second monitoring campaign conducted from the Maidanak observatory has yield time-delay estimates measured by \citet{Vakulik2009, Shimanovskaya2015, Tsvetkova2016}. On the same data set, \citet{Eulaers2011} also tried to estimate the time delays but were unsuccessful. The Mercator and WFI monitoring campaigns are for the first time presented and analyzed in this work.

\subsection{Combining various data sets}\label{sec:finalest}

\begin{figure*}[h!]
\centering
\includegraphics[width=18cm]{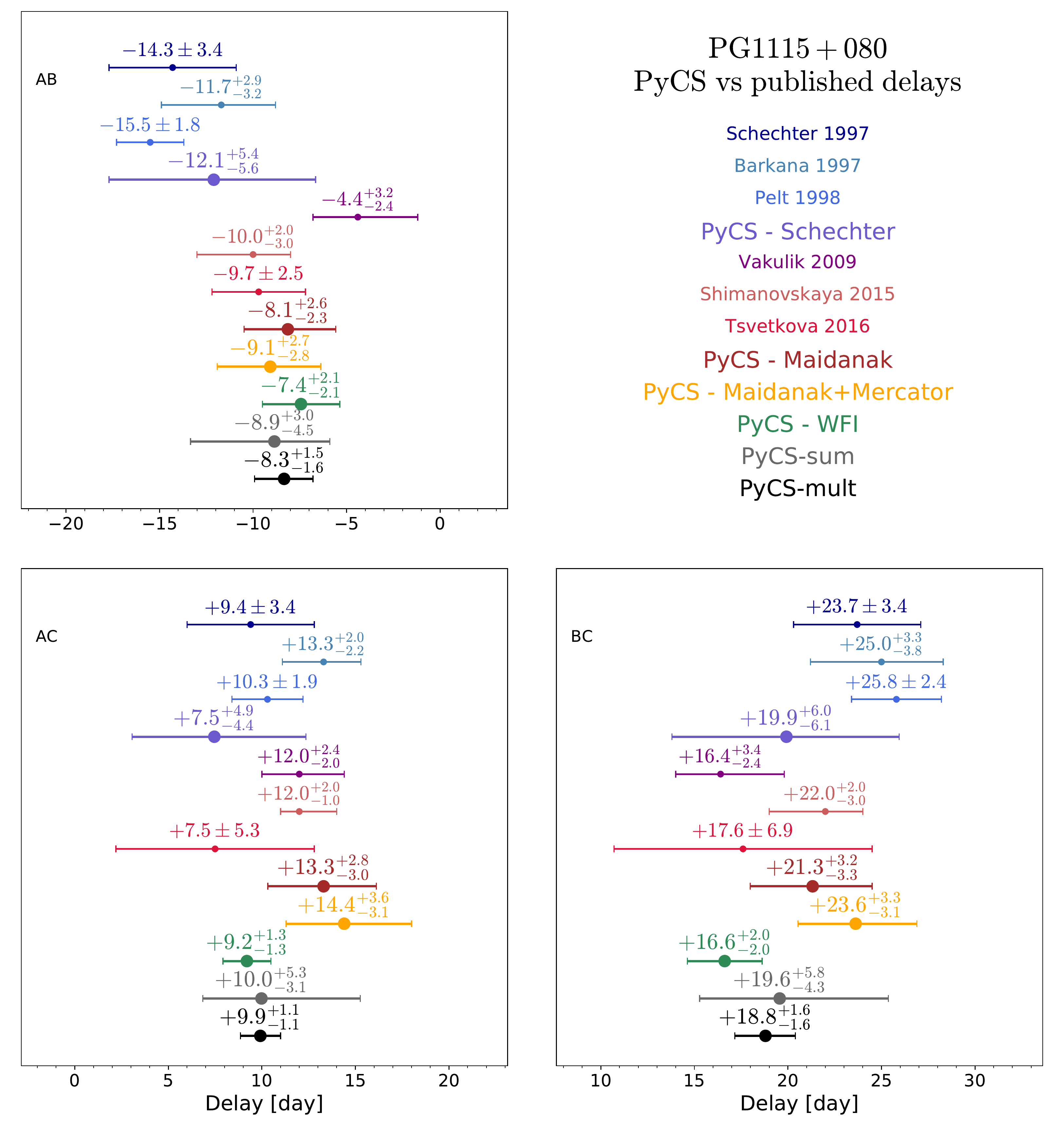}
\caption{Time delays between the images of \obj. Each panel compares the already published values from various authors to our own estimates, obtained using PyCS on the same data 
sets.  The new estimates obtained in this work are labelled ``PyCS'' and are displayed more prominently than the already published estimates. On each panel from top to bottom, the first four estimates are computed 
using the Schechter data set, the four following estimates are computed using 
the Maidanak data set. The last 
two estimates are obtained from two possible combination of our own results on the 
three data sets, either marginalizing over the probability distributions (``PyCS-sum'') or multiplying them (``PyCS-mult''). The quoted mean values and error bars are 
respectively the 50th, 16th and 84th percentiles of the associated time-delay probability 
distributions.}
\label{fig:finalests}
\end{figure*}

The final groups of time-delay estimates for each data set can be combined into a single, final group. There are two ways of performing such a combination. The conservative approach assumes that there might still be shared systematics between the estimates on the three data sets, due to the use of the same curve-shifting techniques. In such a case, the final combined estimates are obtained by marginalizing over the probability distributions corresponding to each estimate. The second approach assumes that the three sets of time-delay estimates are really independent, i.e. that the tension between them (if any) does not results from the curve-shifting techniques used and thus can be combined by multiplying the probability distributions. Asking if the tension hints for unaccounted systematics or can be explained by a statistical fluke can be answered, at least partly, by computing the Bayes Factor $F$ (or evidence ratio) between these two hypothesis. Following \citet{Marshall2006}, we find an evidence of $F_{AB}=56$, $F_{AC}=25$ and $F_{BC}=11$ in favor of the statistical fluke hypothesis. Considering only the most apparent case of tension, i.e. between the BC estimates of WFI and Maidanak+Mercator, we find an evidence of $F_{BC}^{WFI-MM}=1.78$. Without ruling out the possible presence of systematic errors, a Bayes Factor $F>1$ indicates that the considered data sets can be consistently combined into a joint set of time-delay estimates by multiplying the probability distributions.

In Fig.~\ref{fig:finalests}, we present the final combined estimates from the three data sets, where ``PyCS-sum" refers to the marginalization over the three data sets and ``PyCS-mult" refers to the joint set of estimates. From the results of Fig.~\ref{fig:finalests} we can conclude the following:

\begin{enumerate}

\item With the conservative formalism of \pycs, our own time-delay estimates are in comparison less precise than some of the already published estimates, yet always in reasonable agreement.

\item Overall, the WFI data set is the one yielding the most precise time-delay estimates, thanks to the better sampling and photometric precision with respect to the other data sets.

\item The ``PyCS - Maidanak" set of estimates has been obtained by applying the whole analysis pipeline to the Maidanak data only, for the sake of a fair comparison with the literature estimates. It is interesting to note that the addition of the Mercator data to the Maidanak light curves resulted in a slight \emph{decrease} of the overall precision. Such an effect could be explained by the absence of well-defined features in the Mercator light curves, or also by microlensing time delay potentially affecting differently the two monitoring campaigns (see Sec.~\ref{sec:mltd}). The quality of the Mercator data alone is however not sufficient to precisely measure time delays, and the direct comparison with Maidanak is thus not possible. We decided however to use the joint Maidanak+Mercator data set for our final combination, as the addition of extra years of monitoring usually helps constraining the smooth extrinsic variations modeled by the free-knot spline estimator. 

\item The most stringent tension between the individual PyCS estimates is in the BC delay between Maidanak+Mercator and WFI. Using Eq.~\ref{eq:maxtension}, we end up with a tension of $\tau=\sim1.9\sigma$. This tension can result from various factors. First, the Maidanak data reduction has been done using a different pipeline that was not under our control, making it hard to exclude a possible systematic bias in the deconvolution. Second, the timescale of the intrinsic variations observed in the Maidanak+Mercator being longer than in WFI, it is more prone to be degenerate with the extrinsic variations. Third, it could also result from a statistical fluke - a 2 sigma tension has a few percents probability to arise by chance. Last but not least, it could result from microlensing time delay, a systematic error explored in more details in Sec.~\ref{sec:mltd}.

\item The ``PyCS-sum" estimates, although less precise than their ``PyCS-mult" counterpart predict a similar mean value of the time delays. Choosing one or the other for cosmological parameters inference will have an impact on the precision rather than on the accuracy of the results. Being confident that our curve-shifting techniques are sufficiently accurate \citep{Liao2015, Bonvin2016}, we recommend the use of the joint estimates, i.e. the ``PyCS-mult" results.

\end{enumerate}

\section{Effect of the microlensing time delay}\label{sec:mltd}

Not to be confused with the traditional microlensing magnification already implemented in \pycs, a microlensing time delay arises when the accretion disk of the quasar is differently magnified by microlenses (stars or other compact objects) located at the position of the lensed images around the lens galaxy. If the accrection disk is modeled following a lamp-post model \citep{Cackett2007}, temperature variations correlate with luminosity variations. When temperature changes at the center of the accretion disk, it propagates along the disk and generates correlated emission on its way, lagged by the time taken for the impulse to propagate from the center to the edges. Thus, the larger the disk, the longer the lag. In the case of no microlensing, these lagged emissions are order(s) of magnitude fainter than the central emission and are contributing similarly from image to image to the integrated emission. However, for a given magnification pattern, different regions of the accretion disk will be differently magnified, and the lagged contributions will contribute differently to the integrated emission from one lensed image to another. In practice, the accretion disk being far too small to be resolved, light curves of images affected by microlensing time delay are seen shifted in time and skewed with respect to the case of no microlensing, resulting in a biased measurement of the time delays.  

\citet{Tie2017}, who first introduced microlensing time delay, compute its amplitude for the two lensed quasars HE0435-1223 and RXJ1131-1231. They found that the amplitude depends on the size of the accretion disk of the quasar, its orientation relative to the lens and the amount of microlenses at the position of the lensed images. The microlenses and accretion disk are moving with respect to each other, resulting in a time-variable micromagnification of the disk over many years. However, microlensing time delay does not average out over time. Considering the worst cases \citep[see Tab.~2 of][]{Tie2017}, the mean bias is of the order of a day. However, for peculiar geometrical configurations this bias can reach several days. Since the relative motion of the accretion disk and microlenses is slow, such a strong bias can affect the light curves for years. Thus, data sets of short duration like the WFI and Schechter data sets are more likely to be strongly affected, if during this short period of time the quasar happened to lie close to a micro caustic. In data sets with longer baseline such as the Maidanak+Mercator data set, it would be in principle possible to observe a variation of the measured time delays over the years, although in practice the temporal sampling and photometric precision of our light curves are not sufficient to see an effect season by season. 

In the case of \obj\ our three data sets span over two decades, thus if microlensing time delay is at play we should see variations in the measured delays over time. A look at Fig.~\ref{fig:alldelays} shows that the measurements are indeed in slight tension, especially the AC and BC time-delays from the WFI and Maidanak+Mercator data sets - the results from the Schechter data sets being not precise enough to conclude in that regard. Attributing the tension solely to microlensing time delay is certainly wrong, yet it indicates that we have no reason not to consider microlensing time delay as a plausible source of systematic. To address it, we follow the same analysis carried out in \citet{Tie2017} but using microlensing characteristics related to \obj\ instead. We present below the main steps of the analysis and redirect the interested reader to \citet{Tie2017} for more details.

\begin{table}\label{tab:macroinput}
\caption{The $\kappa$, $\gamma$, and $\kappa_{\star}/\kappa$ at each lensed image position from the macro model, based on the modeling in Chen et al. 2018b, in prep.}
\centering
 \begin{tabular}{c c c c}
 Image & $\kappa$ & $\gamma$ & $\kappa_{\star}/\kappa$ \\ [0ex]
 \hline\hline
 $A_{1}$ & 0.424 & 0.491 &0.259\\
 $A_{2}$ & 0.451 & 0.626 &0.263\\
 B & 0.502 & 0.811  &0.331\\ C & 0.356 & 0.315 & 0.203\\
 [0ex] 
 \hline
\end{tabular}
\end{table}

\begin{figure*}[h!]
\centering
\includegraphics[width=18cm]{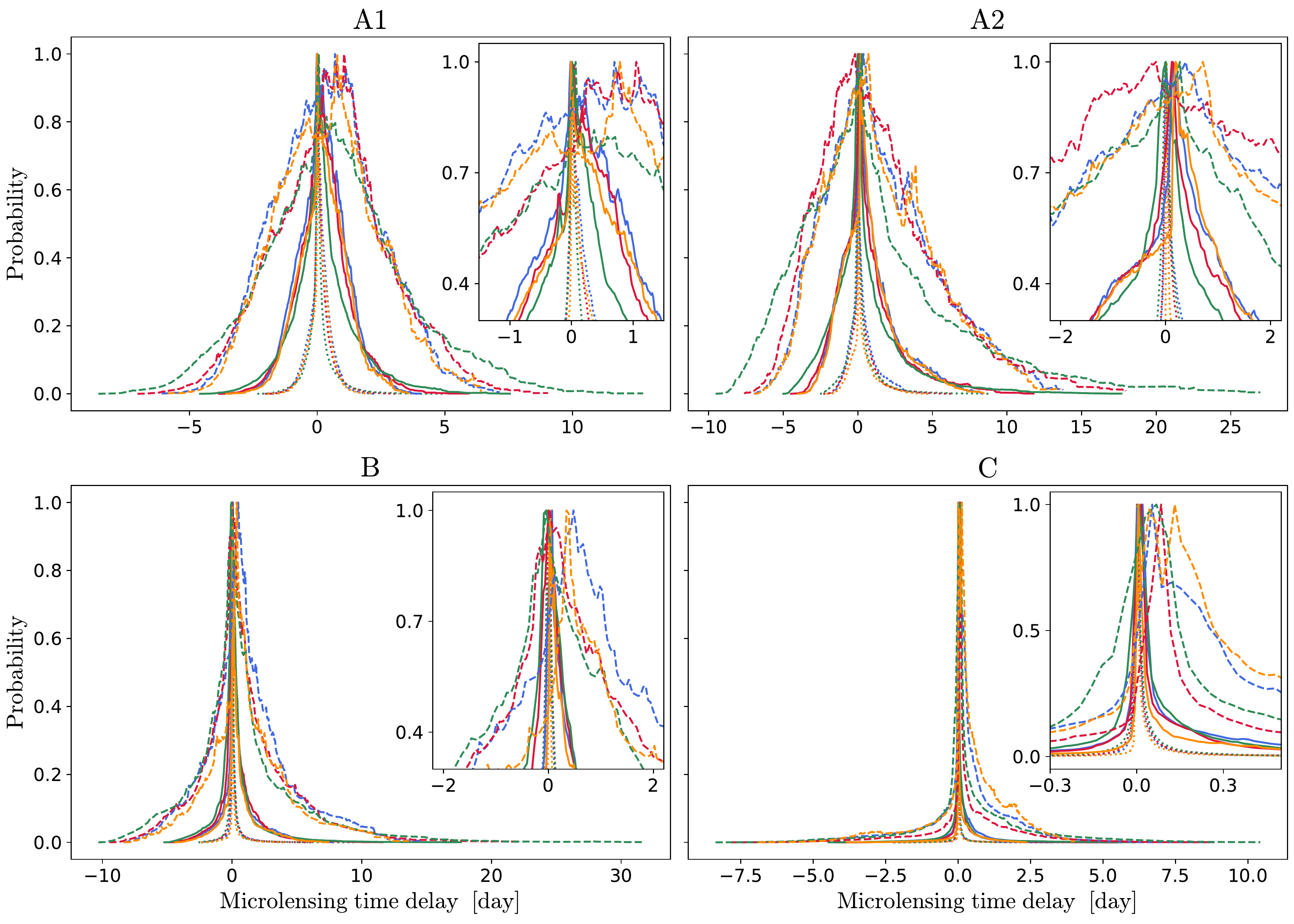}

\vspace{0.5cm}
\begin{tabular}{r c c c c c c c c c c c c c}

  $R_0$ & $i$ & $PA$ & symbol & & A1 & A2 & A & B & C & & AB & AC & BC \\ \hline

\vspace{0.1cm} 0.5 & 0 & 0 & \raisebox{-2.5pt}{\textcolor{darkorange}{\scaleobj{2.2}{\cdot\cdot\cdot}}} &    & $0.09^{+0.36}_{-0.35}$ & $0.12^{+0.92}_{-0.47}$ & $0.14^{+0.49}_{-0.34}$ & $0.05^{+0.39}_{-0.08}$ & $0.01^{+0.11}_{-0.01}$ &    & $-0.07^{+0.55}_{-0.59}$ & $-0.13^{+0.40}_{-0.54}$ & $-0.04^{+0.25}_{-0.46}$ \\

\vspace{0.1cm} 0.5 & 60 & 0 & \raisebox{-2.5pt}{\textcolor{royalblue}{\scaleobj{2.2}{\cdot\cdot\cdot}}} &    & $0.08^{+0.38}_{-0.34}$ & $0.13^{+0.93}_{-0.49}$ & $0.14^{+0.51}_{-0.34}$ & $0.04^{+0.38}_{-0.09}$ & $0.00^{+0.12}_{-0.01}$ &    & $-0.07^{+0.55}_{-0.62}$ & $-0.13^{+0.41}_{-0.56}$ & $-0.04^{+0.25}_{-0.46}$ \\

\vspace{0.1cm} 0.5 & 60 & 45 & \raisebox{-2.5pt}{\textcolor{crimson}{\scaleobj{2.2}{\cdot\cdot\cdot}}} &    & $0.07^{+0.38}_{-0.33}$ & $0.09^{+0.85}_{-0.47}$ & $0.11^{+0.50}_{-0.33}$ & $0.03^{+0.31}_{-0.12}$ & $0.00^{+0.10}_{-0.02}$ &    & $-0.08^{+0.53}_{-0.60}$ & $-0.11^{+0.40}_{-0.55}$ & $-0.03^{+0.24}_{-0.38}$ \\

\vspace{0.1cm} 0.5 & 60 & 90 & \raisebox{-2.5pt}{\textcolor{seagreen}{\scaleobj{2.2}{\cdot\cdot\cdot}}} &    & $0.02^{+0.44}_{-0.31}$ & $0.03^{+0.82}_{-0.52}$ & $0.06^{+0.54}_{-0.37}$ & $0.02^{+0.24}_{-0.19}$ & $0.00^{+0.09}_{-0.04}$ &    & $-0.05^{+0.56}_{-0.67}$ & $-0.05^{+0.45}_{-0.60}$ & $-0.01^{+0.29}_{-0.33}$ \\

\vspace{0.1cm} 1.0 & 0 & 0 & \raisebox{1.0pt}{\textcolor{darkorange}{\scaleobj{1.6}{\rule{0.5cm}{0.8pt}}}} &    & $0.22^{+0.90}_{-0.90}$ & $0.41^{+1.94}_{-1.39}$ & $0.36^{+1.06}_{-0.85}$ & $0.24^{+1.35}_{-0.62}$ & $0.04^{+0.51}_{-0.09}$ &    & $-0.05^{+1.54}_{-1.49}$ & $-0.27^{+1.00}_{-1.24}$ & $-0.17^{+0.96}_{-1.54}$ \\

\vspace{0.1cm} 1.0 & 60 & 0 & \raisebox{1.0pt}{\textcolor{royalblue}{\scaleobj{1.6}{\rule{0.5cm}{0.8pt}}}} &    & $0.21^{+0.92}_{-0.93}$ & $0.38^{+2.02}_{-1.38}$ & $0.34^{+1.12}_{-0.85}$ & $0.21^{+1.44}_{-0.63}$ & $0.04^{+0.46}_{-0.14}$ &    & $-0.03^{+1.60}_{-1.53}$ & $-0.26^{+1.04}_{-1.27}$ & $-0.17^{+0.97}_{-1.58}$ \\

\vspace{0.1cm} 1.0 & 60 & 45 & \raisebox{1.0pt}{\textcolor{crimson}{\scaleobj{1.6}{\rule{0.5cm}{0.8pt}}}} &    & $0.22^{+0.96}_{-0.90}$ & $0.28^{+2.09}_{-1.37}$ & $0.30^{+1.19}_{-0.88}$ & $0.13^{+1.28}_{-0.56}$ & $0.03^{+0.43}_{-0.14}$ &    & $-0.11^{+1.52}_{-1.59}$ & $-0.26^{+1.08}_{-1.34}$ & $-0.10^{+0.99}_{-1.44}$ \\

\vspace{0.1cm} 1.0 & 60 & 90 & \raisebox{1.0pt}{\textcolor{seagreen}{\scaleobj{1.6}{\rule{0.5cm}{0.8pt}}}} &    & $0.13^{+1.19}_{-0.92}$ & $0.09^{+2.36}_{-1.54}$ & $0.19^{+1.45}_{-0.98}$ & $0.05^{+1.10}_{-0.69}$ & $0.03^{+0.38}_{-0.17}$ &    & $-0.12^{+1.64}_{-1.86}$ & $-0.14^{+1.23}_{-1.58}$ & $-0.02^{+1.12}_{-1.31}$ \\

\vspace{0.1cm} 2.0 & 0 & 0 & \raisebox{-2.5pt}{\textcolor{darkorange}{\scaleobj{2.2}{--}}} &    & $0.42^{+2.04}_{-2.02}$ & $0.99^{+4.28}_{-3.00}$ & $0.82^{+2.29}_{-1.85}$ & $0.83^{+3.17}_{-2.25}$ & $0.26^{+1.21}_{-0.75}$ &    & $0.10^{+3.63}_{-3.43}$ & $-0.53^{+2.24}_{-2.66}$ & $-0.57^{+2.68}_{-3.56}$ \\

\vspace{0.1cm} 2.0 & 60 & 0 & \raisebox{-2.5pt}{\textcolor{royalblue}{\scaleobj{2.2}{--}}} &    & $0.38^{+2.09}_{-2.04}$ & $1.00^{+4.40}_{-3.03}$ & $0.80^{+2.34}_{-1.86}$ & $0.80^{+3.35}_{-2.34}$ & $0.22^{+1.25}_{-0.82}$ &    & $0.07^{+3.76}_{-3.52}$ & $-0.59^{+2.29}_{-2.73}$ & $-0.62^{+2.87}_{-3.75}$ \\

\vspace{0.1cm} 2.0 & 60 & 45 & \raisebox{-2.5pt}{\textcolor{crimson}{\scaleobj{2.2}{--}}} &    & $0.60^{+2.16}_{-2.22}$ & $0.74^{+4.57}_{-3.18}$ & $0.76^{+2.49}_{-1.98}$ & $0.53^{+3.69}_{-2.23}$ & $0.15^{+1.50}_{-0.67}$ &    & $-0.09^{+3.96}_{-3.68}$ & $-0.44^{+2.47}_{-2.93}$ & $-0.29^{+2.90}_{-3.96}$ \\

\vspace{0.1cm} 2.0 & 60 & 90 & \raisebox{-2.5pt}{\textcolor{seagreen}{\scaleobj{2.2}{--}}} &    & $0.58^{+2.74}_{-2.48}$ & $0.20^{+5.93}_{-3.73}$ & $0.57^{+3.22}_{-2.36}$ & $0.21^{+3.95}_{-2.44}$ & $0.11^{+1.58}_{-0.72}$ &    & $-0.30^{+4.43}_{-4.50}$ & $-0.34^{+2.96}_{-3.68}$ & $-0.03^{+3.43}_{-4.16}$ \\

\end{tabular}

\caption{Distributions of the excess of microlensing time delay for the four images of \obj. The table below the Figure reports the 16th, 50th and 84th percentiles of the single image distributions as well as the image pair distributions (see text for details) for the various geometrical configurations explored in this work.}
\label{fig:mltd_distrib}
\end{figure*}

The magnification maps for each lensed image are generated using GPU-D \citep{Vernardos2014}, which is a GPU-accelerated implementation of the inverse ray-shooting technique \citep{Wambsganss1992}. We list the microlensing parameters we used in Tab.~\ref{tab:macroinput}. They are based on the lens modeling performed in Chen et al. (2018b, in prep), but we also perform our analysis using the parameters proposed in Tab.~1 of \citet{Morgan2008} for a stellar fraction $f_{M/L}$ of 0.7 and 0.8 and find similar results. We assume a mean mass of the microlenses of $\langle M \rangle = 0.3 M_\odot$ following the Salpeter mass function with a ratio of the upper and lower masses of $r=100$ \citep{Kochanek2004}. Note that our tests showed that the choice of mass function influences little the conclusions below.  
Each map has the size of $20 \langle R_{\rm Ein} \rangle$ with a $8192$-pixel resolution, where

\begin{equation}
\langle R_{\rm Ein} \rangle =\sqrt{\frac{D_{\rm s}D_{\rm ls}}{D_{\rm l}}\frac{4G\langle M \rangle }{c^2}} = 3.618 \times 10^{16} {\rm cm},
\end{equation}

\noindent which depends on the angular diameter distances from the observer to the lens $D_{\rm l}$, the observer to the source $D_{\rm s}$, and the lens to the source $D_{\rm ls}$.

To model the quasar accretion disk, we consider a standard thin disk model \citep{Shakura1973}, which has a radius $R_0 = 1.629\times10^{15}$ cm in the WFI $R_c$ filter (6517.25 \AA) for an Eddington ratio of $L/L_{\rm E}=0.1$ and a radiative efficiency of $\eta=0.1$, given an estimated black hole mass of $1.2\times10^{9} M_\odot$ from \citet{Peng2006}.
Ignoring the inner edge of the disc, in the simple lamp post model of variability the average microlensing time delay can be derived using Eq.10 of \citet{Tie2017}, reproduced here for convenience:

\begin{equation}
\langle\delta t\rangle = \frac{1+z_s}{c}\frac{\int du\ dv\ G(\xi)\ M(u, v)\ R(1+\cos\theta \sin i)}{\int du\ dv\ G(\xi)\ M(u, v)},
\end{equation}

\noindent where $G(\xi)$ is the 1st derivative of the luminosity profile of the disk, $\xi=(R/R_0)^{3/4}$, $M(u, v)$ is the magnification map projected in the source plane and $u,v$ are the observed coordinates in the lens plane \citep[see][for a detailed explanation of the coordinate system]{Tie2017}. $i$ and $\theta$ represent the inclination and position angle of the disk with respect to the source plane, taken as perpendicular to the observer's line of sight. For a given geometrical configuration and accretion disk model, we can thus compute the mean excess of microlensing time delay $\langle \delta t \rangle$ for a given source position and magnification pattern. By varying the magnification pattern, we can infer a distribution of $\langle \delta t \rangle$ for each lensed image.

In this work, we investigate four disc configurations with inclination $i$ and position angle $PA$: i) $i = 0^\circ$, ii)  $i=60^\circ, PA=0^\circ$, iii) $i=60^\circ, PA=45^\circ$, and iv) $i=60^\circ, PA=90^\circ$. Note that the long axis of the tilted disc is perpendicular (parallel) to the caustic structures for $PA = 0^\circ (PA = 90^\circ)$ and corresponds to a face-on disc. We also investigate the effect of decreasing and increasing the source size $R_0$ by a factor of 2. The inferred microlensing time delay distributions, along with their 16th, 50th and 84th percentiles are presented in Fig.~\ref{fig:mltd_distrib}. We have subtracted the contribution due to the lamp post delay of $5.04(1+z_{s})R_0/c$. It corresponds to the excess of time delay that would be present even without any microlensing magnification, that cancels out when measuring time delays between two lensed images.

The mean microlensing time delays from different source configurations follow the trend in \citet{Tie2017}. When the disc is perpendicular to the line of sight the microlensing time delay is longer, the disk size $R_0$ drives the amplitude of the effect, and the median excess of microlensing time delay per lensed image is positive, meaning the effect does not fully average out over time. In the worst case scenario explored in this work, the median shift is of half a day, but can reach several days in a few unlucky cases. Note that contrary to \citet{Tie2017}, we report here the percentile values instead of mean and standard deviation of the distributions. The latter are correct approximations only if the distribution follows a Gaussian profile, which is not necessarily the case (see Fig.~\ref{fig:mltd_distrib}). Depending on the configuration considered, the difference between the mean and 50th percentile can reach a factor of two, the former usually predicting a stronger bias than the latter. It is also interesting to note that the microlensing time delay biases for the C image are much smaller than their A1, A2 and B counterparts, due to the lower fraction of stellar mass $\kappa_{\star}/\kappa$ and lower $\kappa$ at the C image position.

Propagating the microlensing time delay into our time-delay estimates is done the following way. We first compute the microlensing time delay for image A as the mean of the A1 and A2 individual microlensing delays. This is achieved by convolving the A1 and A2 distributions and rescaling the result by a factor 2. Then, we compute the microlensing time delay affecting each pair of images. To do so, in order to take into account that we observe a difference of microlensing time delays between the lensed images, we mirror one of the distribution with respect to zero before convolving them with each other (in other words, we cross-correlate them). The 50th, 16th and 84th percentiles of the distributions for each pair of images are presented in the Table accompanying Fig.~\ref{fig:mltd_distrib}. To propagate these distributions into the time-delay estimates, one would in turn convolve with the time-delay estimate probability distributions of each data set estimated in Sec.~\ref{sec:tds}.

Ultimately, the question that arises is if microlensing time delay should be added to the time-delay measurements or not. As stated earlier, the $1.9\sigma$ tension between the WFI and Maidanak+Mercator data sets, if not uniquely due to microlensing time delay, speaks in favor of it. On the other hand, as mentioned by \citet{Tie2017}, not all quasars are well modeled by the thin-disk model, nor by the lamp-post model of variability. Study of accretion disks with microlensing generally finds larger sources sizes that those predicted by the thin-disk model \citep[see e.g.][and references therein]{Morgan2010, Rojas2014, JimenezVicente2015}, and a similar trend emerges from reverberation mapping studies \citep[e.g.][]{Edelson2015, Lira2015, Fausnaugh2016}, which motivated the exploration of larger source sizes in this section. The microlensing time delay relies on assumptions about astrophysics that are currently hard to verify experimentally. In conclusion, further work is needed to assess if the effect is still present and with which amplitude, for e.g. different accretion disk models. In such cases mitigation strategies could be derived, for example by monitoring the quasar in different bands.

We show that the average microlensing time delay values, albeit different from zero are still small enough not to significantly affect our measured time delays. Presently, we choose not to include the microlensing time delay into our final time delay estimates. All our results are therefore given with error bars that do not include microlensing time delay. However, we present in Sec.~\ref{sec:checks} how the ``PyCS-mult" estimates change when it is taken into account following the formalism presented in this Section. We also redirect the interested readers to Chen et al. (2018a, submitted) for a full account of the microlensing time delay at the lens modeling stage.

\section{Robustness checks}\label{sec:checks}

\begin{figure*}[h!]
\centering
\includegraphics[width=18cm]{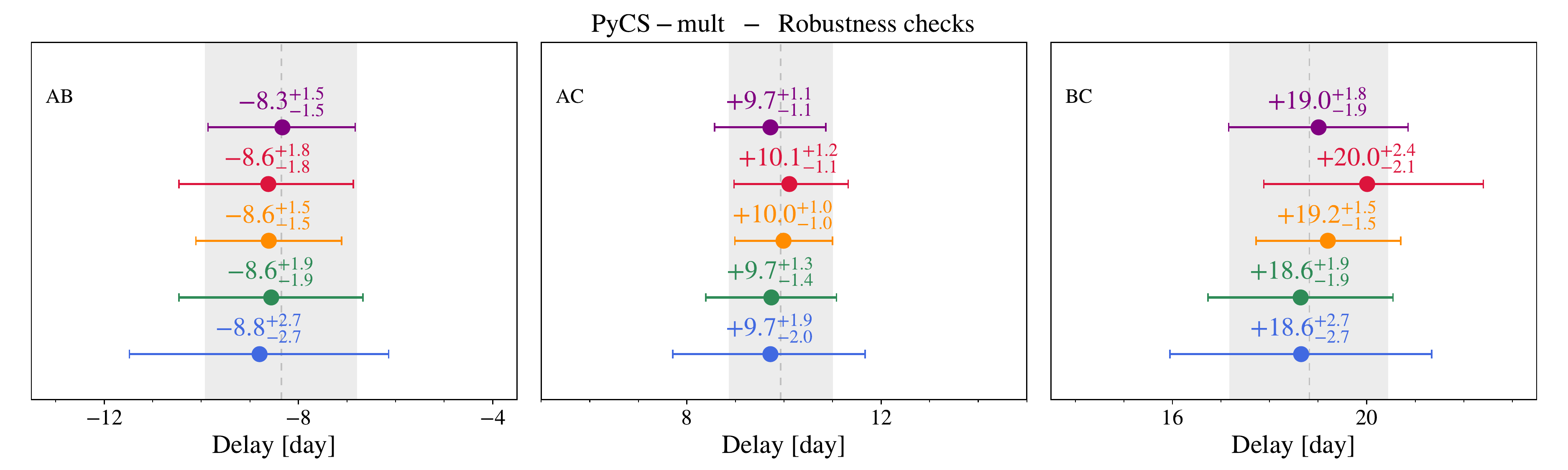}

\begin{tabular}{r  l  r  l  r  l}
  \raisebox{-1.5pt}{\textcolor{purple}{\scaleobj{1.6}{\bullet}}} & WFI alternative microlensing &
 \raisebox{-1.5pt}{\textcolor{crimson}{\scaleobj{1.6}{\bullet}}} & 
  $\tau_{\rm{thresh}}=0\sigma$ &
  \raisebox{-1.5pt}{\textcolor{darkorange}{\scaleobj{1.6}{\bullet}}} & $\tau_{\rm{thresh}}=1\sigma$ \\

 \raisebox{-1.5pt}{\textcolor{seagreen}{\scaleobj{1.6}{\bullet}}} & 
  microlensing time delay with $R=1R_{0}$ &
  \raisebox{-1.5pt}{\textcolor{royalblue}{\scaleobj{1.6}{\bullet}}} & microlensing time delay with $R=2R_{0}$ \\  

\end{tabular}

  \caption{Results of various robustness checks performed in Sec.~\ref{sec:checks}. The fiducial ``PyCS-mult" time-delay estimates from Fig.~\ref{fig:finalests} are reproduced here as shaded gray regions.}

\label{fig:checks}
\end{figure*}

The results presented in Sec.~\ref{sec:finalest} were obtained by marginalizing our curve-shifting techniques over a range of estimators and associated parameters implemented in \pycs. However, not all possible combinations were exhaustively explored and constraining choices were made, e.g. on how the slow extrinsic variations in the free-knot splines technique were handled. In this section, our goal is to assess whether choosing other options beyond those explored in Sec.~\ref{sec:finalest} have a minimal impact on the results. To do so, we use the tension as defined in Eq.\ref{eq:maxtension} to compare the results. In what follows, we refer to the ``PyCS-mult'' time-delay estimates obtained in Sec.~\ref{sec:finalest} as the fiducial estimates to perform the following checks:

\begin{itemize}

\item We explore various ranges of estimator parameters for the regression difference technique. As highlighted by \citet{Steinhardt2018}, it is not possible to know a priori if a Gaussian process regression has converged to its best possible solution, hence the importance of exploring a large range of possible combinations. Limiting ourselves to only five choices of estimator parmeter combinations is purely artificial and future improvements of this curve-shifting technique should include a way to go beyond this limitation, e.g. by using priors with adaptive constraints on the estimator parameters. In the present case, we tested the regression difference estimator against extreme values of the estimator parameters, well outside the range used for the fiducial estimates. This resulted in similar median values but with much larger error bars. The tension stayed always below 0.5$\sigma$ with the fiducial estimates.

\item We vary the microlensing model used in the Maidanak+Mercator data set analysis. Testing against various microlensing models is a good way to assess whether the chosen model is biased by, e.g. features of the intrinsic variations being accounted for in the extrinsic variations, or vice-versa. The chosen extrinsic spline initial knot step $\eta_{\rm{ml}}$ was increased to 360 and 500 days, either keeping the minimal spacing between the knots at 100 days or increasing it to 180 and 250 days, respectively. In both cases, both the precision and accuracy were similar to the results presented in Sec.~\ref{sec:tds}, with a tension with the fiducial results always below 0.5$\sigma$. Assuming there is no microlensing has a much stronger effect, shifting the mean measured time-delays by up to several days. Yet, it is impossible to properly stack the three light curves across all five seasons without allowing for microlensing variability. For this reason, we believe that the results without microlensing should not be used. We avoid decreasing $\eta_{\rm{ml}}$ below 100 as in such a regime, intrinsic and extrinsic variations would become degenerate and bias the outcome.

\item We vary the microlensing model used in the Schechter and WFI data set analysis. Similar to the previous point, we want to assess whether the model chosen in the fiducial analysis has any degeneracies between intrinsic and extrinsic features in the light curves. We avoid adding more than one knot to the extrinsic splines since it would make the intrinsic and extrinsic variations degenerate. Instead, we explore an alternative solution by giving more freedom to the central knot, letting its position on the time axis slide up to 50 days closer to both ends of the extrinsic splines. Doing so adds degeneracy between the intrinsic and extrinsic splines, yet remains an interesting robustness test to perform. For the Schechter data set, this alternative microlensing model marginally affects the results, notably because the precision is relatively low. Thus, the maximum tension with the fiducial results is only 0.15$\sigma$. For the WFI data set, since the precision is much better the tension between the alternative microlensing and fiducial results goes up to $\sim0.6\sigma$ for the free-knot splines technique with $\eta=20$ and $\eta=30$. Since this exceeds the fiducial threshold $\tau_{\rm{thresh}}=0.5\sigma$ used for the combination of time-delay estimates in Sec.~\ref{sec:tds}, we performed the whole analysis using this new microlensing model for WFI instead of the fiducial one. Interestingly, the regression difference results computed using the modified microlensing in the generative model for mock light curves remained very close to their fiducial counterparts, with a maximum tension of $0.2\sigma$. Whereas the fiducial regression difference and free-knot splines time-delay estimates on WFI were in excellent agreement, as it can be seen on Fig.~\ref{fig:alldelays}, this is less true for the modified microlensing results. The fiducial combined WFI results are thus more precise than the modified microlensing ones, the resulting tension between the two being smaller than 0.5$\sigma$. The impact on the final joint combination is also barely noticeable, with a maximum tension of $\sim0.1$, and is represented in purple on Fig.~\ref{fig:checks}.

\item The sigma threshold value used when combining the sets of time-delay estimates for a given curve-shifting technique and data set, initially chosen at $\tau_{\rm{thresh}}=0.5\sigma$, has no motivations other than accounting for the variance of the estimates for a given estimator. We want to make sure that this arbitrary choice has no strong effect on the outcome. We consider here two extreme cases: with $\tau_{\rm{thresh}}=1.0\sigma$, the pipeline simply picks the most precise estimates per panel in Fig.~\ref{fig:alldelays}. With $\tau_{\rm{thresh}}=0\sigma$, all estimates of Fig.~\ref{fig:alldelays} are combined together. Note that both cases are not very reasonable choices: neither neglecting estimates in tension with the fiducial estimates, nor including known unprecise estimates is correct. However, doing so provides valuable information on the robustness of our final time-delay estimates. The results of this process is presented in Fig.~\ref{fig:checks}. The maximum tension with the fiducial value is of $\sim0.2$ for $\tau_{\rm{thresh}}=1\sigma$ and $\sim0.4$ for $\tau_{\rm{thresh}}=0\sigma$. Both measurements are represented in red and orange on Fig.~\ref{fig:checks}.

\item We include the microlensing time delay following the formalism presented in Sec.~\ref{sec:mltd}, i.e. convolving the microlensing time delay distribution to the time delay measurement error distribution. This is done on the ``PyCS-Schechter", ``PyCS-Maidanak+Mercator" and ``PyCS-WFI" set of time-delay estimates individually, before combining them into a single set. We present in Fig.~\ref{fig:checks} the result from two configurations: we fix the inclination angle and position at zero, and use source sizes of $1R_0$ and $2R_0$. The impact on the final result remains moderate, with a maximum change in accuracy of $3.5\%$ ($6\%$) and a relative decrease of precision, computed through Eq.~\ref{eq:precision}, of $21.1\%$ ($71.1\%$) for a source size of $1R_0$ ($2R_0$). Note that an alternative approach to include microlensing time delay at the lens modeling stage instead of the measurement stage is presented in Chen et al. (2018a, submitted) and illustrated with the microlensing time delay results on \obj\ presented in Sec.~\ref{sec:mltd} of this paper.

\end{itemize}

In this section we tested some specific effects that we thought could affect our results, but as shown in Fig.~\ref{fig:checks} these remain of low impact. When compared to the fiducial ``PyCS-mult'' measurements, the tension always stayed below 0.5$\sigma$, thus assessing the robustness of our results.

\section{Conclusions}

In this paper, we present the light curves of the lensed quasar \obj\ after one season of monitoring at the \tel. We expand this monitoring campaign with the already published data from various telescopes in the years 1996-1997 \citep{Schechter1997} as well as data taken between 2004 and 2006 at the Maidanak telescope in Uzbekistan \citep{Tsvetkova2010}. We complement the latter data set with three monitoring seasons at the Mercator telescope taken from 2005 to 2008.

We present individual measurements of the time delays on these data sets using \pycs, a curve-shifting toolbox developed over the years in the COSMOGRAIL collaboration. We notably include in our results estimation of the microlensing time-delay, following the formalism introduced in \citet{Tie2017} as well as a marginalization strategy over the choice of curve-shifting techniques and optimizer parameters in \pycs. Our results are in agreement with previous estimates from the literature. The time-delay estimates obtained using the \tel\ monitoring data are the most precise estimates published so far. This demonstrates along with \citet{Courbin2017} how \emph{quasi daily observations over a single season at very high signal-to-noise ratio can surpass long-term monitoring carried out less frequently over many seasons}.

By combining our measurements on all the data sets, we obtain values for the time delays (without including the microlensing time delay) of $\Delta t(AB) = 8.3^{+1.5}_{-1.6}$ days (18.7\% precision), $\Delta t(AC) = 9.9^{+1.1}_{-1.1}$ days (11.1\%) and $\Delta t(BC) = 18.8^{+1.6}_{-1.6}$ days (8.5\%). Our results are robust against how extrinsic intensity variations from microlensing are modeled and how individual set of estimates are combined.

We compute the impact of microlensing time delay for various source parameters. Explicitly accounting for it in the time-delay measurements results in a loss of precision that depend mostly on the chosen size of the accretion disk. We decided not to include it in our final time-delay estimates, as i) it relies on astrophysical assumptions that are not yet proven to be true (the accretion disk follow a lamp-post model of variability, and is well modeled by a thin-disk model), ii) there is no clear evidence of microlensing time delay in our data and iii) a more efficient formalism to handle microlensing time delay at the lens modeling stage is presented in a companion paper (Chen et al. 2018a, submitted).

Cosmological inference with \obj\ will be carried out in a dedicated paper (Chen et al. 2018b, in prep.) using AO imaging from the Keck telescope. With two of the three time delays measured around the 10\% precision level, \obj\ will be very useful for cosmography when included in a joint analysis of a larger sample of lensed quasars \citep{Treu2016}. Ongoing large-sky surveys (e.g. STRIDES, KiDS, CFIS) and future ones (e.g. LSST,  Euclid)
will drastically increase the number of known lens systems \citep[e.g.][]{Oguri2010}. In such a context, dedicated monitoring telescopes that can yield robust time-delay estimates in a single monitoring season will be crucial.

High-cadence monitoring at the \tel\ started in October 2016. Since then, six different lensed quasars have already been monitored for a full season already and three more are currently being monitored on a daily basis. Among these, four targets have been recently discovered and never been monitored before. This work follows the presentation of the time delays of the lensed quasar DES J0408-5354 \citep{Courbin2017}, and represents the second installment of a series of time-delay measurements from high-cadence monitoring soon to be extended.

\begin{acknowledgements}
The authors would like to thank R. Gredel for his help in setting up the programme at the \tel. This work is supported by the Swiss National Fundation. This research made use of Astropy, a community-developed core Python package for Astronomy \citep{Astropy2013, Astropy2018} and the 2D graphics environment Matplotlib \citep{Hunter2007}. K.R. acknowledge support from PhD fellowship FIB-UV 2015/2016 and Becas de Doctorado Nacional CONICYT 2017 and thanks the LSSTC Data Science Fellowship Program, her time as a Fellow has benefited this work. M.T. acknowledges support by the DFG grant Hi 1495/2-1. G.~C.-F.~C. acknowledges support from the Ministry of Education in Taiwan via Government Scholarship to Study Abroad (GSSA). D.~C.-Y.~Chao and S.~H.~Suyu gratefully acknowledge the support from the Max Planck Society through the Max Planck Research Group for S.~H.~Suyu. T. A. acknowledges support by the Ministry for the Economy, Development, and Tourism's Programa Inicativa Cient\'{i}fica Milenio through grant IC 12009, awarded to The Millennium Institute of Astrophysics (MAS).
\end{acknowledgements}

\bibliographystyle{aa}
\bibliography{paper}

\clearpage

\end{document}